\documentclass[aps, prd, twocolumn, floatfix, nofootinbib, superscriptaddress]{revtex4-1}

\usepackage{amsmath,amsfonts,amssymb,bm}
\usepackage{dsfont}
\usepackage{graphicx}
\usepackage{color}
\usepackage{soul}
\usepackage{slashed}

\usepackage[mathscr,scaled=1.15]{urwchancal}
\DeclareFontFamily{OT1}{pzc}{}
\DeclareFontShape{OT1}{pzc}{m}{it}%
{<-> s * [1.15] pzcmi7t}{}
\DeclareMathAlphabet{\mathpzc}{OT1}{pzc}{m}{it}

\definecolor{purple}{rgb}{0.5,0,0.5}
\definecolor{blue}{rgb}{0.0,0,0.9}
\definecolor{prdblue}{rgb}{0.133,0.118,0.498}
\usepackage[colorlinks=true, pdfstartview=FitV, linkcolor=prdblue, citecolor= prdblue, urlcolor=prdblue]{hyperref}

\begin{document}

\title{Locating the Gribov horizon}

\author{Fei Gao}
\affiliation{Department of Physics and State Key Laboratory of Nuclear Physics and Technology, Peking University, Beijing 100871, China}

\author{Si-Xue~Qin}
\affiliation{Department of Physics, Chongqing University, Chongqing 401331, P.R. China}

\author{Craig D.~Roberts}
\affiliation{Physics Division, Argonne National Laboratory, Argonne, Illinois 60439, USA}

\author{Jose Rodr\'iguez-Quintero}
\affiliation{Departamento de F\'isica Aplicada, Facultad de Ciencias Experimentales, Universidad de Huelva, Huelva E-21071, Spain}

\date{15 June 2017}

\begin{abstract}
We explore whether a tree-level expression for the gluon two-point function, supposed to express effects of an horizon term introduced to eliminate the Gribov ambiguity, is consistent with the propagator obtained in simulations of lattice-regularised quantum chromodynamics (QCD).  In doing so, we insist that the gluon two-point function obey constraints that ensure a minimal level of consistency with parton-like behaviour at ultraviolet momenta.  In consequence, we are led to a position which supports a conjecture that the gluon mass and horizon scale are equivalent emergent mass-scales, each with a value of roughly $0.5\,$GeV; and wherefrom it appears plausible that the dynamical generation of a running gluon mass may alone be sufficient to remove the Gribov ambiguity.
\end{abstract}

\maketitle




\noindent\textbf{1.$\;$Introduction}.
When quantising continuum chromodynamics, a gauge fixing condition must be imposed upon the gluon fields.  Except in particular cases \cite{Gribov:1977wm, Konetschny:1979mw, Zwanziger:1989mf, Baulieu:1981ec, Chan:1985kf, Vink:1992ys, Bowman:2002bm, Williams:2003du, Zhang:2004gv}, that cannot be completed consistently without adding ghost fields to the Lagrangian \cite{Faddeev:1967fc}.  The classical theory's gauge invariance is then replaced by BRST symmetry \cite{Becchi:1975nq, Tyutin:1975qk}, which can be used in perturbation theory to prove, \emph{e.g}.\ renormalisability of quantum chromodynamics (QCD).

Typically, however, the auxiliary condition meant to select a unique element from each class of equivalent configurations (a gauge field orbit) is nonperturbatively inadequate \cite{Gribov:1977wm, Konetschny:1979mw, Zwanziger:1989mf, Baulieu:1981ec, Chan:1985kf, Vink:1992ys, Bowman:2002bm, Williams:2003du, Zhang:2004gv}.  An unknown (probably infinite) number of configurations remain, each related to the identified element by a nonperturbative gauge transformation, and all contributing equally to the functional integral that should define the theory.  This impedes a rigorous mathematical formulation of QCD; and hence the domain of gauge field integration must be restricted further \cite{Gribov:1977wm, Zwanziger:1989mf}.

Contemporary efforts to realise a gauge fixing procedure that selects a unique configuration from each gauge field orbit are described in Ref.\,\cite{Vandersickel:2012tz}.   The analysis is typically undertaken in Landau gauge, because, \emph{e.g}.: it is a linear covariant gauge; a fixed point of the renormalisation group; and readily implemented in lattice-QCD.
Prominent amongst associated schemes is a modification of the standard QCD action to include an ``horizon term'':
\begin{equation}
\label{horizon}
\gamma \, \int d^4 x\, h(x) \,,\,
%
\end{equation}
%
$h(x)  = g^2 f^{abc} A_\mu^b(x) [{\mathpzc M}^{-1}]^{ad}(x,x) f^{dec} A_\mu^e(x)$,
where $g$ is the coupling, $\{A_\mu^a\}$ represents the gluons, and
\begin{equation}
{\mathpzc M}^{ab}(x,y) = 
[- \partial^2 \delta^{ab} + \partial_\mu f^{abc} A_\mu^c(x)]\delta^4(x-y)
\end{equation}
is the Landau gauge Faddeev-Popov operator.  The scale $\gamma$ is fixed via the ``horizon condition'':
$\langle h[\gamma] \rangle = d (N^2-1)$,
where $d (N^2-1)$, $d=4$, $N=3$, is the number of components of the gluon field, and the expectation value indicates gauge-field integration in the presence of $\gamma$.  (Issues of BRST (a-)symmetry, renormalisability, etc., of such an action are canvassed elsewhere, \emph{e.g}.\ Refs.\,\cite{Dudal:2010fq, Vandersickel:2012tz}.)

The procedure just described ensures that only those solutions of the Landau-gauge auxiliary condition which produce non-negative values for the Faddeev-Popov determinant ($Det\, {\mathpzc M} \geq 0$) contribute in the gauge-field integration, \emph{i.e}.\ it restricts the integration to those Landau-gauge configurations which lie within the so-called first Gribov region, $\Omega$, whose boundary, $\delta \Omega$, the ``Gribov horizon'', is defined by Landau-gauge configurations for which $Det \,{\mathpzc M} =0$.
The importance of these requirements is clear: $Det {\mathpzc M} \neq 0$ is necessary to ensure the existence of a unique solution of the gauge fixing condition and $Det {\mathpzc M} \geq 0$ is required if the determinant is to be represented using ghost fields.
The first Gribov region has other significant properties  \cite{Vandersickel:2012tz}: $\Omega$ contains the trivial $A_\mu^0 \equiv 0$ configuration, so a connection with perturbation theory is maintained; and it is intersected by each gauge orbit at least once, convex, and compact.  It follows that all gauge orbits are represented by configurations within $\Omega$ and any set of Gribov copies within $\Omega$ is bounded.

The procedure summarised here refines the original scheme \cite{Faddeev:1967fc}; and its implementation capitalises on a finding \cite{Capri:2005dy} that one may equivalently define the first Gribov region to be the set of relative minima of the functional
\begin{equation}
F_A[U] = \tfrac{1}{2} \int d^4x\, [A_\mu^a(x)]_U \, [A_\mu^a(x)]_U\,,
\end{equation}
where $[A_\mu^a]_U$ is a gauge transformation of the field $A_\mu^a$ and the minimisation proceeds by choosing those configurations on each orbit which minimise this norm.  Plainly, there can be more than one relative minima; and the set of gauge-equivalent minima identifies the Gribov copies tied to a given configuration.  From this perspective, the Landau-gauge ambiguity is resolved if the gauge field integration is restricted to those configurations for which $F_A[U]$ is an absolute minimum.  The domain of such configurations defines the fundamental modular region, $\Lambda \subset \Omega$, and quantisation of QCD may then properly be achieved by integrating only over $A_\mu^a \in \Lambda$.  However, an issue remains \cite{Williams:2003du}: no practical, local scheme has yet been devised to achieve this restriction in continuum QCD.

Our discussion highlights that the Gribov horizon, $\delta\Omega$, might play an important role in rigorously defining the scope of gauge sector interactions.  Its existence is: imposed via $\gamma > 0 $ in Eq.\,\eqref{horizon}, whose value is dynamically determined; and known to have the potential to modify the infrared (IR) behaviour of the gluon two-point Schwinger function \cite{Zwanziger:1988jt}.  Hence, writing $g^2 \gamma = m_\gamma^4$, one may identify $m_\gamma$ as an interaction-induced mass scale whose value characterises the location of the Gribov horizon.

Implemented as described, the gauge-fixed action is non-local because it involves ${\mathpzc M}^{-1}$.  At the cost of introducing additional fields, an equivalent local action can be derived, yielding a tree-level gluon propagator for a theory whose gauge fields all lie within $\Omega$ \cite{Gribov:1977wm, Zwanziger:1989mf}:
\begin{equation}
\label{gluongamma}
D^\gamma_{\mu\nu}(k) = T_{\mu\nu}(k) \mathpzc{D}^\gamma(k^2)\,,\;
\mathpzc{D}^\gamma(k^2) = \frac{k^2}{k^4 + 2 N g^2 \gamma}\,,
\end{equation}
$T_{\mu\nu}(k) = \delta_{\mu\nu}-k_\mu k_\nu/k^2$.  It is now plausible to suppose that Eq.\,\eqref{gluongamma} expresses the dominant IR features of the gauge-fixed gluon propagator and hence the scheme employed could be validated through comparison with Landau-gauge results for this Schwinger function obtained using lattice-QCD (lQCD).  The most striking feature of Eq.\,\eqref{gluongamma} is that $\mathpzc{D}^\gamma(k^2)\to 0$ as $k^2 \to 0$ and it is now clear that such behaviour is \emph{not} found in QCD.  Instead, the gluon sector is characterised by a nonperturbatively-generated IR mass-scale, which ensures the gluon dressing function is nonzero and finite at $k^2=0$ \cite{Aguilar:2006gr, Cucchieri:2007md, Cucchieri:2007rg, Aguilar:2008xm, Dudal:2008sp, Brodsky:2008be, Bogolubsky:2009dc, Oliveira:2010xc, Aguilar:2011ux, Boucaud:2011ug, Aguilar:2012rz, Binosi:2014aea, Aguilar:2015bud, Cyrol:2016tym, Binosi:2016nme}.  One must therefore conclude that, notwithstanding its strengths, the scheme of Refs.\,\cite{Gribov:1977wm, Zwanziger:1989mf} is incomplete.

A modification to the gauge-fixing scheme of Refs.\,\cite{Gribov:1977wm, Zwanziger:1989mf} is canvassed in Refs.\,\cite{Dudal:2007cw, Dudal:2008sp}.  It admits the possibility that the ghost fields used to localise the horizon term develop a nonzero dimension-two condensate, whose presence further modifies the gluon propagator \cite{Dudal:2007cw, Dudal:2008sp}:
\begin{equation}
\label{overlineD}
\mathpzc{D}^\gamma(k^2) \to
\overline{\mathpzc{D}}(k^2)
= \frac{k^2 + M^2}{k^4 + k^2 \mathpzc{m}^2 + \lambda^4}\,,
\end{equation}
where
$\lambda^4 = 2 N g^2 \gamma - \mu^2 M^2$,
$\mathpzc{m}^2=M^2-\mu^2$,
with $\mu^2\propto \langle A_\mu^a A_\mu^a\rangle$ and $M^2$ related to the ghost-field condensate, both computed within an hadronic medium \cite{Brodsky:2012ku}.  Following this method, which supports a nonzero value for the gluon propagator in the far-IR, one can obtain fair agreement with lQCD results for the gluon two-point function \cite{Dudal:2010tf, Cucchieri:2011ig, Dudal:2012zx}.
Herein, we revisit this issue, using lattice simulations to constrain the parameters in Eq.\,\eqref{overlineD} after imposing novel constraints; and this enables us to develop new insights into their implications and meaning.

\smallskip

\noindent\textbf{2.$\;$Partonic Constraints}.
Owing to asymptotic freedom, and notwithstanding confinement and dynamical chiral symmetry breaking, the scalar functions characterising any one of the two-point Schwinger functions associated with QCD's elementary excitations can be defined as positive on $k^2\in [0,\infty)$ and must then be convex-down and fall no faster than $1/k^2$, with modest logarithmic corrections, on
${\mathpzc P} = \{ k^2\,|\, k^2> k_{\mathpzc P}^2, \mbox{\rm for~some}\; k_{\mathpzc P}^2 > \Lambda_{\rm QCD}^2\}$.
Denoting such a function by ${\mathpzc S}(k^2)$, then equally, as may be shown using the operator product expansion \cite{Wilson:1969zs}, $\exists \tau_{\mathpzc P}< 1/\Lambda_{\rm QCD}$ such that $\sigma(\tau)$, the configuration space dual of ${\mathpzc S}(k^2)$, is convex-down on
${\mathpzc P}_\tau = \{ \tau\,|\, 0 <  \tau < \tau_{\mathpzc P}\}$.
It is commonly assumed that such duals are straightforwardly related by $\cos$-transform \cite{Hawes:1993ef}.  This is a strong claim because one cannot prove nonperturbatively that the transform yields the same function obtained by computing the correlator directly in configuration space.  Nevertheless, it is currently impossible to do better; and under this assumption, the impact of these convexity requirements (partonic constraints) on $\overline{\mathpzc{D}}(k^2)$ in Eq.\,\eqref{overlineD} may be elucidated by considering the associated $\cos$-transform:
\begin{align}
\Delta(\tau) &= 
 \nonumber
\Delta_{\mathpzc P}(\tau)
\left[
 (1+M^2/\lambda^2) s_{\varphi/2} \cos(\tau \lambda s_{\varphi/2} ) \right. \\
 & \left. \quad \quad
-(1-M^2/\lambda^2) c_{\varphi/2} \sin(\tau \lambda s_{\varphi/2} )
\right]\,,
\label{Deltatau}
\end{align}
where $\Delta_{\mathpzc P}(\tau)=\exp(-\tau \lambda c_{\varphi/2})/(2 \lambda s_\varphi)$,
$s_\varphi = [1-c_\varphi^2]^{1/2}$, $s_{\varphi/2} = [1-c_{\varphi/2}^2]^{1/2}$,
$c_\varphi=\cos\varphi = \mathpzc{m}^2/(2\lambda^2)$,
$c_{\varphi/2} = \cos\tfrac{1}{2}\varphi  = [\tfrac{1}{2} + \mathpzc{m}^2/(4\lambda^2)]^{1/2}$.
%
%

The Schwinger function in Eq.\,\eqref{overlineD} is that of a massive excitation, in which case a propagator consistent with the partonic constraint would behave as $\exp(- {\rm mass} \times \tau)$.  In order to ensure this, it is \emph{sufficient} to demand
\begin{equation}
\label{sufficient}
M^2=\lambda^2\,,\; \lambda^2 \geq \tfrac{1}{3} \mu^2 \quad (\mbox{sufficient}) \,,
\end{equation}
for then $\left.\Delta(\tau) \right|_{M=\lambda} = \Delta_{\mathpzc P}(\tau)
[ 1 + \mbox{O}(\tau^2 \lambda^2 s_{\varphi/2}^2)]$.
%

Of course, a \emph{necessary} condition for convexity of $\Delta(\tau)$ on $\tau\simeq 0$ is that $\Delta^{\prime\prime}(\tau)$ be positive on this domain:
\begin{align}
\left. \frac{d^2}{d\tau^2} \Delta(\tau) \right|_{\tau =0} = \frac{1}{4 \lambda c_{\varphi/2}} [ 1 - \mu^2/\lambda^2]\,,
\end{align}
which is positive so long as
\begin{equation}
\label{necessary}
\mu^2\leq \lambda^2 \quad \mbox{(necessary)} \,.
\end{equation}

\smallskip

\noindent\textbf{3.$\;$Intepreting Lattice-QCD Results}.
We now explore the impact of the partonic constraints in Eqs.\,\eqref{sufficient}, \eqref{necessary} on the description of lQCD results for the gluon two-point function using Eq.\,\eqref{overlineD}.

To proceed, we assume that at some renormalisation scale, $\zeta_{\rm GZ}$, $\overline{\mathpzc D}(k^2;\zeta_{\rm GZ}^2):=\overline{\mathpzc D}(k^2)$ in Eq.\,\eqref{overlineD} is a valid representation of that part of the full gluon two-point function which is essentially nonperturbative.
There is then a domain of IR momenta, $k^2\in [0,k_0^2]$, with $k_0$ to be determined, whereupon Eq.\,\eqref{overlineD} should be capable of describing the lQCD propagator, $D_{\#}$, computed at a known renormalisation scale, $\zeta_\#$, after they are both evolved to a common point $\zeta_0 \sim 1\,$GeV, typical of hadron physics.
On $k^2>k_0^2$, the usual logarithms and anomalous dimensions will be generated by renormalisation.  Since they are absent from Eq.\,\eqref{overlineD}, this formula must thereupon fail.

The gluon two-point function is multiplicatively renormalisable, so this perspective can be expressed thus:
\begin{subequations}
\label{allfit}
\begin{align}
\label{fitparams}
\forall k^2\in [0,k_0^2]\!: & \;\overline{\mathpzc D}(k^2;\zeta_0^2)   = D_{\#}(k^2;\zeta_0^2) \,,\\
\overline{\mathpzc D}(k^2;\zeta_0^2)  & = {\mathpzc z}_0(\zeta_0^2;\zeta_{\rm GZ}^2) \overline{\mathpzc D}(k^2;\zeta_{\rm GZ}^2)\,,  \\
 D_{\#}(k^2;\zeta_0^2) & = D_{\#}(k^2;\zeta_\#^2)  / [ \zeta_0^2 D_{\#}(\zeta_0^2;\zeta_\#^2)]  \,,
\end{align}
\end{subequations}
where $\mathpzc{z}_0$, $k_0$ are fit parameters.  We will judge the interpretation reasonable so long as $\mathpzc{z}_0 \sim 1$, $k_0\sim 1\,{\rm GeV} $, in which event it is natural to identify $\zeta_0 = k_0$.
There are five parameters: $\mathpzc{z}_0$, $k_0=\zeta_0$, and $M$, $\mu$, $\lambda$ in Eq.\,\eqref{overlineD}.  They are determined simultaneously by minimising the rms relative-difference between the two sides of Eq.\,\eqref{fitparams} for a given value of $k_0^2$, then optimising $k_0^2\in [0,k_{\rm max}^2]$ by seeking that value which produces the global minimum for this rms relative-difference, where $k_{\rm max}$ is the largest momentum at which lQCD results are available.

\begin{table}[t]
\caption{
\emph{Panel A}.  Analysis of large-volume quenched lQCD results \cite{Bogolubsky:2009dc} ($\zeta_\#=4.3\,{\rm GeV}$, $k_{\rm max}=4.5\,$GeV) using the two-point function in Eq.\,\eqref{overlineD}.
Row~1: unconstrained fit using simulation results on the entire available domain, $k_0=k_{\rm max}$.
Row~2: unconstrained fit on an IR domain.
Row~3: fit respecting the parton-sufficient  condition, Eq.\,\eqref{sufficient}.
Row~4: fit respecting the parton-necessary condition, Eq.\,\eqref{necessary}, imposing the upper bound.
Row~5: fit respecting Eq.\,\eqref{necessary} and requiring $m_\gamma = m_g$ (see Sec.\,5).
\emph{Panel B}. As Panel\,A, but for unquenched results $(N_f=4)$ \cite{Ayala:2012pb}.
\emph{Panel C}. $k_{\rm ip}^2$ -- position of the inflection point in $\overline{\mathpzc D}(k^2;\zeta_0^2)$ when computed using the coefficients listed in the rows above, \emph{e.g}.\ A1 means Panel\,A, Row 1; $\tau_{\rm z}$ --  location of the first zero in the associated $\Delta(\tau)$; $\tau_{\rm F}$ -- related parton-persistence or fragmentation length.
(All dimensioned quantities in GeV, except $\tau_{\rm z}$, $\tau_{\rm F}$, in fm.)
\label{latticefits}
}
\begin{tabular*}
{\hsize}
{
l|@{\extracolsep{0ptplus1fil}}
c|@{\extracolsep{0ptplus1fil}}
c|@{\extracolsep{0ptplus1fil}}
c|@{\extracolsep{0ptplus1fil}}
c|@{\extracolsep{0ptplus1fil}}
c|@{\extracolsep{0ptplus1fil}}
c|@{\extracolsep{0ptplus1fil}}
c@{\extracolsep{0ptplus1fil}}}\hline
(A) $\phantom{un}$quenched & $k_0$ & $\zeta_0$ & $\lambda$ & $M$  & ${\mathpzc z}_0$ & $M/\lambda$ & $\mu/\lambda$  \\\hline
unconstrained  & 4.5 & 1.1 & 0.84 & 2.10  & 0.43 & 2.49 & 2.33 \\
  & $\zeta_0$ & 1.1 & 0.72 & 1.09 & 0.75 & 1.50 & 1.27  \\\hline
sufficient & $\zeta_0$ & 1.0  & 0.59  & 0.59 & 1.04 & 1 & 0.45  \\
necessary &  $\zeta_0$ & 1.0 & 0.68 & 0.88 & 0.84 & 1.29 & 1  \\\hline
nec.\,$+m_\gamma=m_g$ &  $\zeta_0$ & 1.0 & 0.67 & 0.84 & 0.87 & 1.26 & 0.94  \\\hline
\end{tabular*}

\medskip

\begin{tabular*}
{\hsize}
{
l|@{\extracolsep{0ptplus1fil}}
c|@{\extracolsep{0ptplus1fil}}
c|@{\extracolsep{0ptplus1fil}}
c|@{\extracolsep{0ptplus1fil}}
c|@{\extracolsep{0ptplus1fil}}
c|@{\extracolsep{0ptplus1fil}}
c|@{\extracolsep{0ptplus1fil}}
c@{\extracolsep{0ptplus1fil}}}\hline
(B) unquenched  & $k_0$ & $\zeta_0$ & $\lambda$ & $M$ & ${\mathpzc z}_0$ & $M/\lambda$ & $\mu/\lambda$  \\\hline
unconstrained  & 4.0 & 1.1 & 1.01 & 2.38 & 0.59 & 2.36 & 1.91 \\
  & $\zeta_0$ & 1.1 & 0.95 & 1.92 & 0.65 & 2.02 & 1.64  \\\hline
sufficient & $\zeta_0$ & 1.0 & 0.65 & $0.65$ & 1.20 & 1 & 0.42\,i  \\
necessary & $\zeta_0$ & 1.0 & 0.87 & 1.34 & 0.93 & 1.54 & 1  \\\hline
nec.\,$+m_\gamma=m_g$ &  $\zeta_0$ & 1.0 & 0.80 & 1.07 & 0.89 & 1.34 & 0.71  \\\hline
\end{tabular*}

\medskip

\begin{tabular*}
{\hsize}
{
l|@{\extracolsep{0ptplus1fil}}
c|@{\extracolsep{0ptplus1fil}}
c|@{\extracolsep{0ptplus1fil}}
c|@{\extracolsep{0ptplus1fil}}
c|@{\extracolsep{0ptplus1fil}}
c|@{\extracolsep{0ptplus1fil}}
c|@{\extracolsep{0ptplus1fil}}
c|@{\extracolsep{0ptplus1fil}}
c|@{\extracolsep{0ptplus1fil}}
c|@{\extracolsep{0ptplus1fil}}
c@{\extracolsep{0ptplus1fil}}}\hline
(C) & A1 & A2 & A3 & A4 & A5 & B1 & B2 & B3 & B4 & B5\\\hline
$k_{\rm ip}$ & 0.36 & 0.46 & 0.48 & 0.48 & 0.48 & & 0.30 & 0.42 & 0.40 & 0.14\\
$\tau_{\rm z}$ & 1.04 & 0.99 & 0.96 & 0.97 & 0.97 & & 1.35 & 1.05 & 1.33 & 1.36 \\\hline
$\tau_{\rm F}$ &   &   & 0.67 & 0.81 & 0.80 & &   & 0.66 & 1.17 & 1.02 \\\hline
\end{tabular*}

\end{table}

\begin{figure}[t]
\centerline{\includegraphics[width=0.4\textwidth]{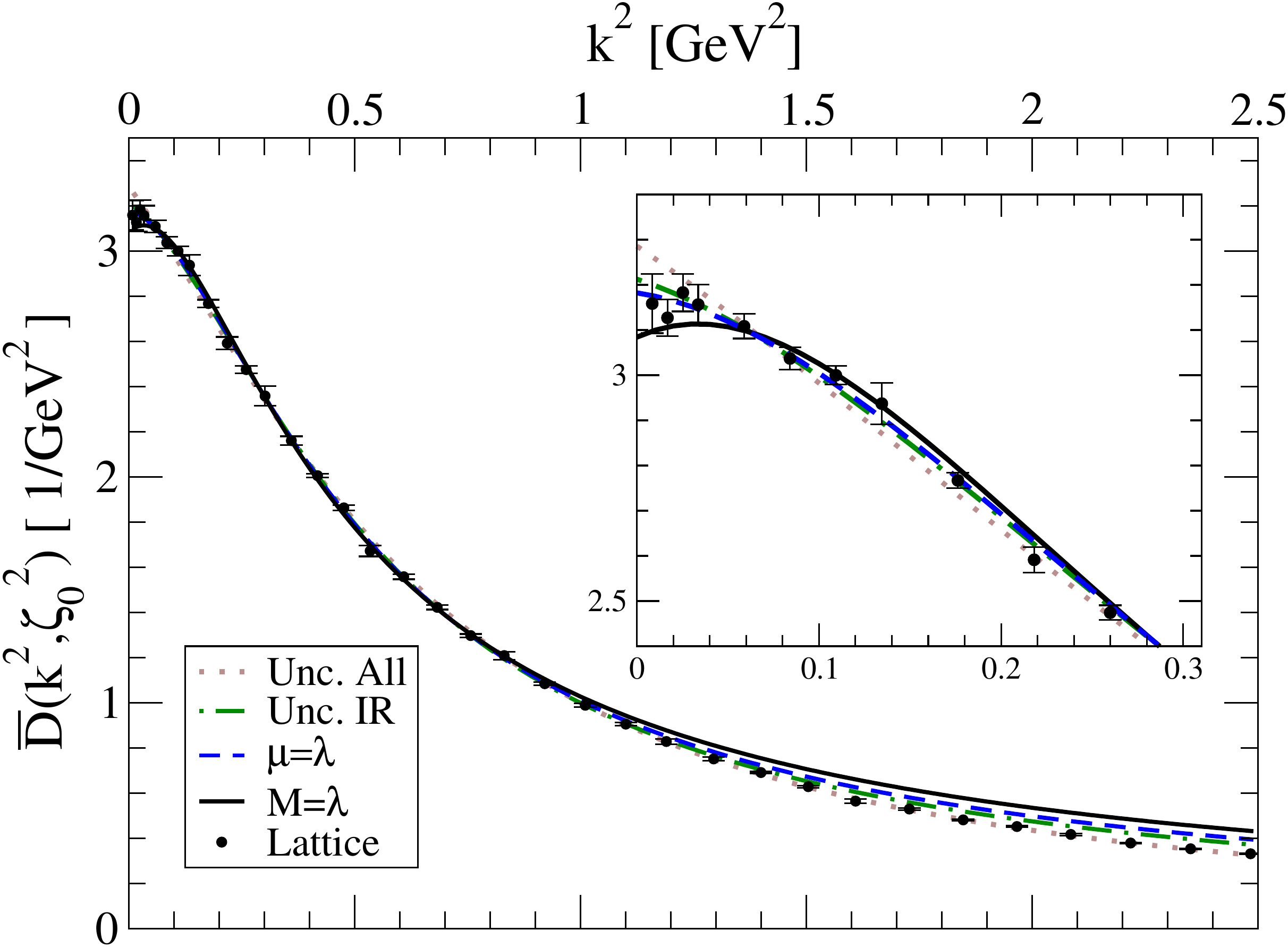}}

\centerline{\includegraphics[width=0.4\textwidth]{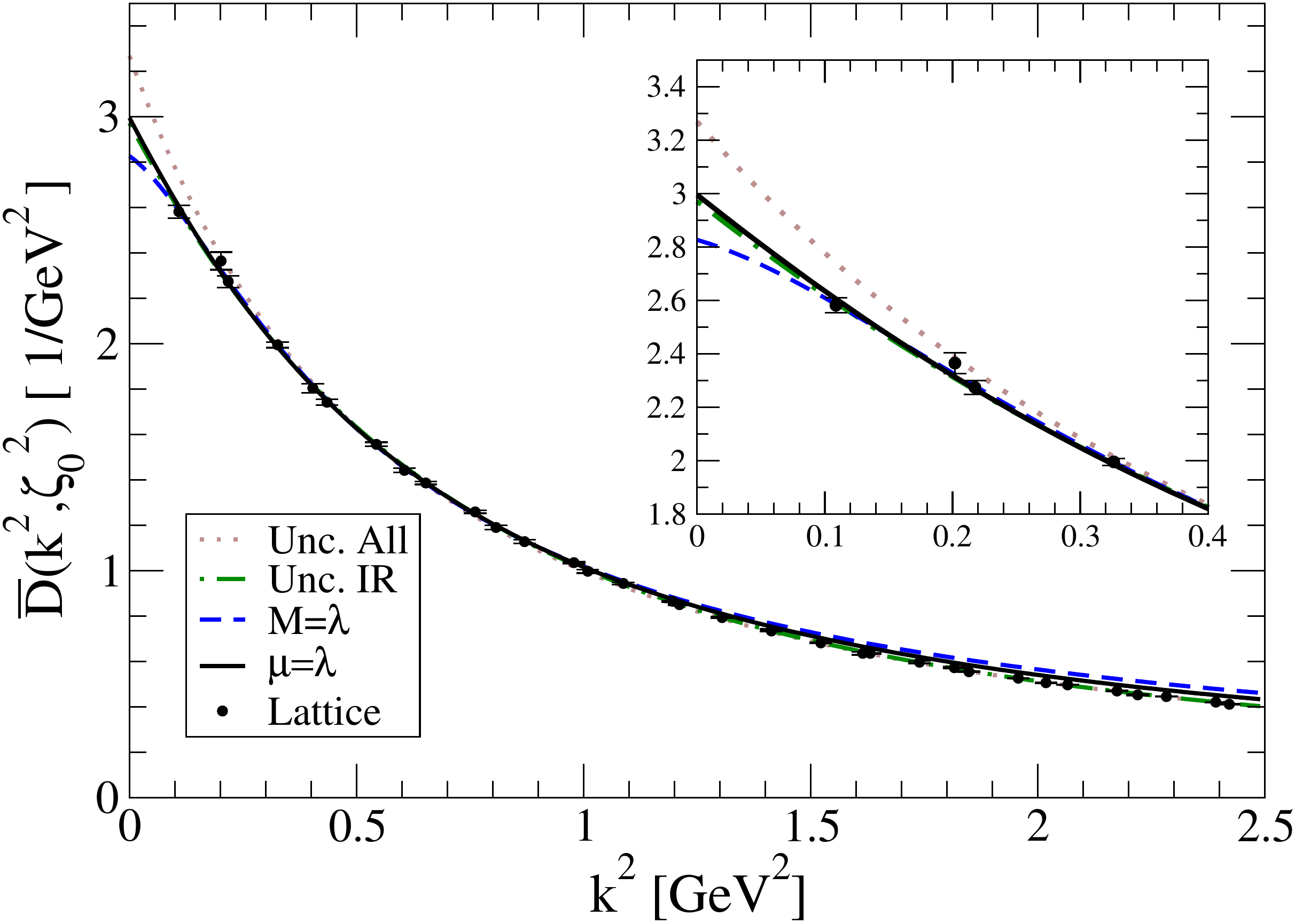}}
\caption{\label{FitQuenched}
\emph{Upper panel - (A).}
Fit to quenched lattice results \cite{Bogolubsky:2009dc}, determined as described in connection with
Eqs.\,\eqref{allfit}:
dotted (brown) curve -- Row~1 of Table~\ref{latticefits}A,  unconstrained, unbounded fit;
dot-dashed (green) curve -- Row~2, Table~\ref{latticefits}A,  unconstrained, bounded fit;
dashed (blue) curve -- Row~3, satisfying the sufficient condition for partonic behaviour, Eq.\,\eqref{sufficient}; and
solid (black) curve -- Row~4, necessary condition, Eq.\,\eqref{necessary}.
\emph{Lower panel - (B)}.  As upper panel, but for unquenched results $(N_f=4)$ \cite{Ayala:2012pb}.
(The curves and points associated with the unconstrained fits have been rescaled by $(1/1.1)^2$.  This eliminates an offset from the constrained results owing to the small difference in optimal scales: $\zeta_0=1.1\,$GeV \emph{cf}.\ $\zeta_0=1.0\,$GeV.)
}
\end{figure}

We first analyse the large-volume quenched simulations in Ref.\,\cite{Bogolubsky:2009dc} ($\zeta_\# = 4.3\,$GeV, $k_{\rm max}=4.5\,$GeV), wherewith our procedure yields the results in Table~\ref{latticefits}A.  For this purpose the $64^4$ and $80^4$ lattices are indistinguishable \cite{Dudal:2012zx}: we use the latter.
Row\,1 reports the coefficients that achieve a best fit on the entire domain of available lattice momenta, \emph{i.e}.\ $k_0:=k_{\rm max}$, unvaried, with $\zeta_0$ fixed at the value found to produce the global minimum when $k_0$ is optimised, \emph{viz}.\ determined in producing Row~2.  The fit's quality is apparent in Fig.\,\ref{FitQuenched}A: the ultraviolet (UV) behaviour is represented well at the cost of a poorer description of the IR.  This is unsurprising, given the preponderance of UV lattice results in the sample domain; and, consequently, ${\mathpzc z}_0$ deviates greatly from unity, indicating that on $k^2\in [0,k_{\rm max}^2]$ the lQCD output possesses material perturbative contributions, which cannot be captured by Eq.\,\eqref{overlineD}.
The fit nevertheless exhibits an IR inflection point [at $k_{\rm ip}^2=(0.36\,{\rm GeV})^2$], signalling that the spectral function associated with this propagator is not positive-definite.  Such behaviour is widely interpreted as an indicator of confinement \cite{Munczek:1983dx, Stingl:1985hx, Krein:1990sf, Burden:1991gd, Hawes:1993ef, Maris:1994ux, Bhagwat:2002tx, Roberts:2007ji, Bashir:2009fv, Ayala:2012pb, Bashir:2013zha, Qin:2013ufa, Lowdon:2015fig, Lucha:2016vte}.  It is expressed in $\Delta(\tau)$ of Eq.\,\eqref{Deltatau} via a non-terminating series of zeros, with the first located at $\tau_{\rm z} = 1.04\,$fm: at this scale, even the most tenuous connection with partonic behaviour is lost.  (Such inflection points and zeros are listed in Table~\ref{latticefits}C.)

Row\,2 in Table~\ref{latticefits}A lists the coefficients obtained when fitting the quenched results on an IR domain with upper bound $k_0$, as described in connection with Eqs.\,\eqref{allfit}.  Here, ${\mathpzc z}_0$ is closer to unity, so an interpretation using Eq.\,\eqref{overlineD} is more credible.  The result is the dot-dashed (green) curve in Fig.\,\ref{FitQuenched}A.  Here, $k_{\rm ip}^2=(0.46\,{\rm GeV})^2$ and the first zero in $\Delta(\tau)$ lies at $\tau_{\rm z}=0.99\,$fm.  Notably, however, neither the sufficient nor necessary condition for partonic behaviour, Eqs.\,\eqref{sufficient}, \eqref{necessary}, is satisfied; and inspection reveals that $\Delta(\tau)$ has an inflection point at $\tau=0.078\,$fm.  Such a striking nonperturbative effect located deep in the UV is in direct conflict with perturbation theory.  Hence, this fit, too, should be rejected as unrealistic.

It is thus appropriate now to describe the fit to lattice results that respects the sufficient condition for parton-consistent behaviour, Eq.\,\eqref{sufficient}: parameters in Row~3 of Table~\ref{latticefits}A and dashed (blue) curve in Fig.\,\ref{FitQuenched}A.  In this instance: ${\mathpzc z}_0$ is close to unity; $\overline{\mathpzc D}(k^2;\zeta_0^2)$ has a global maximum at $k^2=(0.18\,{\rm GeV})^2$ and an inflection point at $k_{\rm ip}^2=(0.48\,{\rm GeV})^2$.
The appearance of a global maximum at $k^2>0$ is, perhaps, unexpected, but neither continuum nor lattice analyses of QCD's gauge sector can exclude this possibility.  In fact, contributions from massless-ghost loops in the gluon vacuum polarisation may produce just such an effect \cite{Aguilar:2013vaa, Binosi:2016xxu}.  It is noteworthy that constraining the behaviour of a two-point function on $\tau \simeq 0$, \emph{viz}.\ a far-UV domain, inaccessible to lattice simulations, enables extraction of more reliable information about the behaviour of the function at IR momenta, a domain within which lattice results are concentrated.

The associated $\Delta(\tau)$ is convex on a domain extending beyond its first zero, at $\tau_{\rm z}=0.96\,$fm.  In this instance one may also ask after the persistence of partonic behaviour, in which connection we define a \emph{hadronisation length} as that scale, $\tau_{\rm H}$, whereat $\Delta(\tau_{\rm H})/\Delta_{\mathpzc P}(\tau_{\rm H})= \tfrac{1}{2}$, \emph{i.e}.\ the configuration-space Schwinger function deviates from a partonic propagator by $\geq 50$\%.  Here, $\tau_{\rm H}=0.67\,$fm.

\begin{table}[t]
\caption{
Dimension-two condensate, $m_{A^2}$, Eq.\,\eqref{g2A2};
gluon mass-scale, $m_g=\lambda^2/M$;
and Gribov horizon parameter, $m_\gamma$, Eq.\,\eqref{GHparameter},
inferred by fitting lattice results using Eq.\,\eqref{overlineD} subject to the parton constraints in Eqs.\,\eqref{sufficient}, \eqref{necessary}, and also subject to Eq.\,\eqref{necessary} plus $m_\gamma = m_g$ (see Sec.\,5).  The resolving scale is $\zeta_0=1\,$GeV.
($\Lambda_{\rm QCD}^{{\rm MOM}(N_f=0)}=0.425(+15)(-9)$ and $\Lambda_{\rm QCD}^{{\rm MOM}(N_f=4)}=0.560(\pm 30)$ \cite{Boucaud:2008gn, Blossier:2012ef, Blossier:2013ioa}.  All dimensioned quantities in GeV.)
\label{GribovHorizon}
}
\begin{tabular*}
{\hsize}
{
c|@{\extracolsep{0ptplus1fil}}
c|@{\extracolsep{0ptplus1fil}}
c|@{\extracolsep{0ptplus1fil}}
c|@{\extracolsep{0ptplus1fil}}
c|@{\extracolsep{0ptplus1fil}}
c|@{\extracolsep{0ptplus1fil}}
c|@{\extracolsep{0ptplus1fil}}
c|@{\extracolsep{0ptplus1fil}}
c@{\extracolsep{0ptplus1fil}}}\hline
 \multicolumn{1}{c|}{$\,$} & \multicolumn{3}{c|}{sufficient}& \multicolumn{3}{c|}{necessary} & \multicolumn{2}{c}{nec.\,$+m_\gamma=m_g$} \\\hline
 $N_f$ & $m_{A^2}$ & $m_g\,$& $m_\gamma\,$ & $m_{A^2}$ & $m_g\,$& $m_\gamma\,$ & $m_{A^2}$ & $m_\gamma=m_g$\\\hline
  0 & $0.86\,\phantom{i}$ & $0.59$ & $0.39$ & 2.22 & $0.53$ & $0.56$ & 2.05 & 0.53 \\
  4 & $0.89\,i$ & 0.65 & $0.40$ & 2.84 & 0.56 & $0.75$ & 1.85 & 0.60 \\\hline
\end{tabular*}
\end{table}

The fit also yields a value for the ``$\langle A^2\rangle$" in-hadron condensate, appearing in the operator product expansion of the gluon two-point function \cite{Boucaud:2001st, Dudal:2008sp, Dudal:2010tf, Boucaud:2011ug, Cucchieri:2011ig, Blossier:2013ioa}:
\begin{equation}
\label{g2A2}
g^2 \langle A^2\rangle = \tfrac{32}{3} \mu^2 =: m_{A^2}^2\,.  
\end{equation}
Phenomenologically: $m_{A^2} \sim (1-3 \,{\rm GeV})^2$; but this fit gives $(0.86\,{\rm GeV})^2$, as reported in Table~\ref{GribovHorizon}.

Row~4 in Table~\ref{latticefits}A specifies the Eq.\,\eqref{necessary}-consistent fit, which is the solid (black) curve in Fig.\,\ref{FitQuenched}A: little visually distinguishes the results obtained from the sufficient and necessary parton constraints.  Here: ${\mathpzc z}_0$ is close to unity; $\overline{\mathpzc D}(k^2;\zeta_0^2)$ has an inflection point at $k_{\rm ip}^2=(0.48\,{\rm GeV})^2$; and $\Delta(\tau)$ is convex on a domain that extends beyond its first zero, located at $\tau_{\rm z}=0.97\,$fm, with $\tau_{H}=0.81\,$fm.

We now turn to unquenched $(N_f=4)$ results \cite{Ayala:2012pb} ($\zeta_\# = 4.3\,$GeV, volume $48^3\times 96$).  Our procedure yields the results listed in Table~\ref{latticefits}B.  As evident in Fig.\,\ref{FitQuenched}B, there is a paucity of unquenched results at IR momenta, one consequence of which is a need to bound the fitting window in order achieve any reasonable description.  Thus, the first row in Table~\ref{latticefits}B lists the coefficients required to achieve a best fit on $k^2 \leq k_{\rm max}^2$, $k_{\rm max}=4.0\,$GeV, with $\zeta_0$ fixed at the value found to produce the global minimum when $k_0$ is optimised, \emph{viz}.\ determined in producing Row~2.  The fit is poor at IR momenta (see the dotted (brown) curve in Fig.\,\ref{FitQuenched}B), owing to the scarcity of results, which also entails that the fit does not describe a manifestly confined excitation: $\overline{\mathpzc D}(k^2;\zeta_0^2)$ has no inflection point at spacelike momenta and $\Delta(\tau\geq 0)>0$.  Furthermore, as in the analogous quenched case, ${\mathpzc z}_0$ is far from unity owing to the presence of material perturbative contributions to the lQCD results on this fitting domain.

Row~2 in Table~\ref{latticefits}B specifies the fit to unquenched results on $k^2\in [0,\zeta_0^2]$, $\zeta_0=1.1\,$GeV. Again, ${\mathpzc z}_0$ is quite different from unity.  The result: dot-dashed (green) curve in Fig.\,\ref{FitQuenched}B, possesses a first-order inflection point, \emph{viz}.\ the first $k^2$-derivative of $\overline{\mathpzc D}(k^2;\zeta_0^2)$ exhibits an inflection point at $k_{\rm ip}^2=(0.30\,{\rm GeV})^2$, and yields a form for $\Delta(\tau)$ whose first zero lies at $\tau_{\rm z}=1.35\,$fm.  Notably, as with the analogous quenched case, neither Eq.\,\eqref{sufficient} nor Eq.\,\eqref{necessary} is satisfied; and inspection reveals that $\Delta(\tau)$ has an inflection point at $\tau=0.10\,$fm.  Hence, this fit is unrealistic.

The impact of the sufficient condition, Eq.\,\eqref{sufficient}, on fitting the unquenched propagator is expressed in Row~3, Table~\ref{latticefits}B.  Shown as the dashed (blue) curve in Fig.\,\ref{FitQuenched}B, it displays an inflection point at $k_{\rm ip}^2=(0.42\,{\rm GeV})^2$.  Computed therefrom, $\Delta(\tau)$ is convex on a domain that extends beyond its first zero, at $\tau_{\rm z}=1.05\,$fm, with $\tau_{\rm F}=0.66\,$fm.  Once again, imposing a physical constraint on the behaviour of a two-point function at far-UV momenta has enabled extraction of more reliable information about its IR behaviour.  Notwithstanding these features, $|{\mathpzc z}_0-1|$ is still too large for us to be confident that the unquenched results are consistent with Eq.\,\eqref{overlineD}, especially because $\mu^2<1$ and hence $\langle A_\mu^a A_\mu^a \rangle$ has the wrong sign when compared with contemporary phenomenology.

Imposing Eq.\,\eqref{necessary} (necessary) when analysing the unquenched propagator yields Row~4 of Table~\ref{latticefits}B and the two-point function depicted by the solid (black) curve in Fig.\,\ref{FitQuenched}B.  There is a clear difference between the impact of the sufficient and necessary conditions.  The necessary condition ensures ${\mathpzc z}_0 \sim 1$ and the fit is characterised by $k_{\rm ip}^2=(0.40\,{\rm GeV})^2$, $\tau_{\rm z}=1.33\,$fm, and $\tau_{\rm F}=1.17\,$fm.

\smallskip

\noindent\textbf{4.$\;$Gribov Horizon Parameter}.
Using the coefficients in Table~\ref{latticefits}, one can compute the Gribov horizon scale:
\begin{equation}
\label{GHparameter}
m_\gamma^4 := g^2 \gamma =\tfrac{1}{2N} \left[ \lambda^4 +  \mu^2 M^2 \right]\,,
\end{equation}
with the results listed in Table~\ref{GribovHorizon} and compared with the dressed-gluon mass-scale inferred from the same ensembles: $m_g = \lambda^2/M$.

The pattern of the results in Table~\ref{GribovHorizon} is readily explained.
For instance, compared with the unquenched values, the sizes of the quenched results are typically smaller owing to the paucity of unquenched IR data, which leads that fit to a focus on UV momenta and thus, usually, increased magnitudes for $\lambda$, $M$, $\mu$.
%
Likewise, the sizes obtained with the parton-sufficient constraint, Eq.\,\eqref{sufficient}, are normally smaller than those found with the necessary constraint, Eq.\,\eqref{necessary}, because the latter also forces larger magnitudes for $\lambda$, $M$, $\mu$.

Focusing now on the Gribov parameter itself, the sufficient condition entails $m_\gamma < m_g$, whereas the necessary condition favours $m_\gamma \gtrsim m_g$.  These outcomes are embedded in Eq.\,\eqref{GHparameter}, \emph{e.g}.\ implementing Eq.\,\eqref{sufficient}, one has
\begin{align}
m_\gamma^4 - m_g^4
%
& \stackrel{N=3}{= } \tfrac{\lambda^2}{6} \left[ \mu^2 - 5 \lambda^2\right]\,,
\end{align}
which is negative for all values of $\mu^2$ consistent with Eq.\,\eqref{sufficient}.  On the other hand, enforcing Eq.\,\eqref{necessary}:
\begin{align}
m_\gamma^4 - m_g^4 & \stackrel{N=3}{\leq} \tfrac{\lambda^2}{6} \left[ \lambda^2 + M^2 - 6\tfrac{ \lambda^6}{M^4}\right]\,.
\end{align}
The rhs is positive $\forall M > 1.24\,\lambda$, a condition satisfied by all fits except, naturally, those obtained using Eq.\,\eqref{sufficient}.
Which constraint, then, is more realistic?  The value of $m_{\rm A^2}$ suggests the necessary condition is better aligned with phenomenology.  (The unconstrained fits produce unrealistically large values of $m_{\rm A^2}$ and are thus excluded.)  Notably, for $\mu^2>0$ the sufficient condition always produces a global maximum in $\overline{\mathpzc D}(k^2;\zeta_0^2)$ at $k^2>0$.  Hence, a preference for the necessary condition places an upper bound on the strength of contributions from massless-ghost loops to the gluon vacuum polarisation.

\smallskip

\noindent\textbf{5.$\;$Conclusion}.
Reflecting upon our results, consider that in the context of QCD augmented by the horizon term, Eq.\,\eqref{horizon}, there are three scenarios.

\begin{figure}[t]
\centerline{\includegraphics[width=0.4\textwidth]{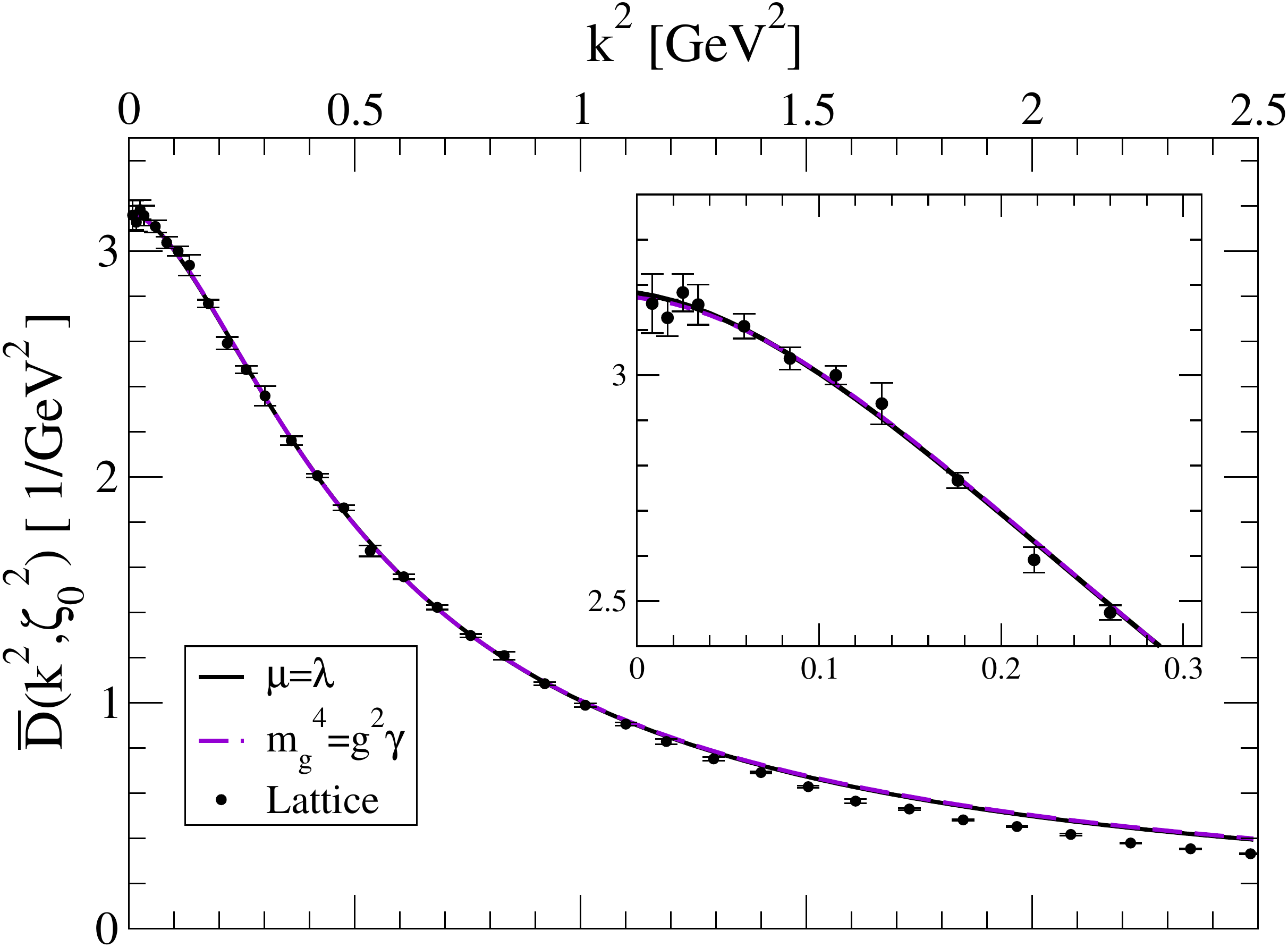}}

\centerline{\includegraphics[width=0.4\textwidth]{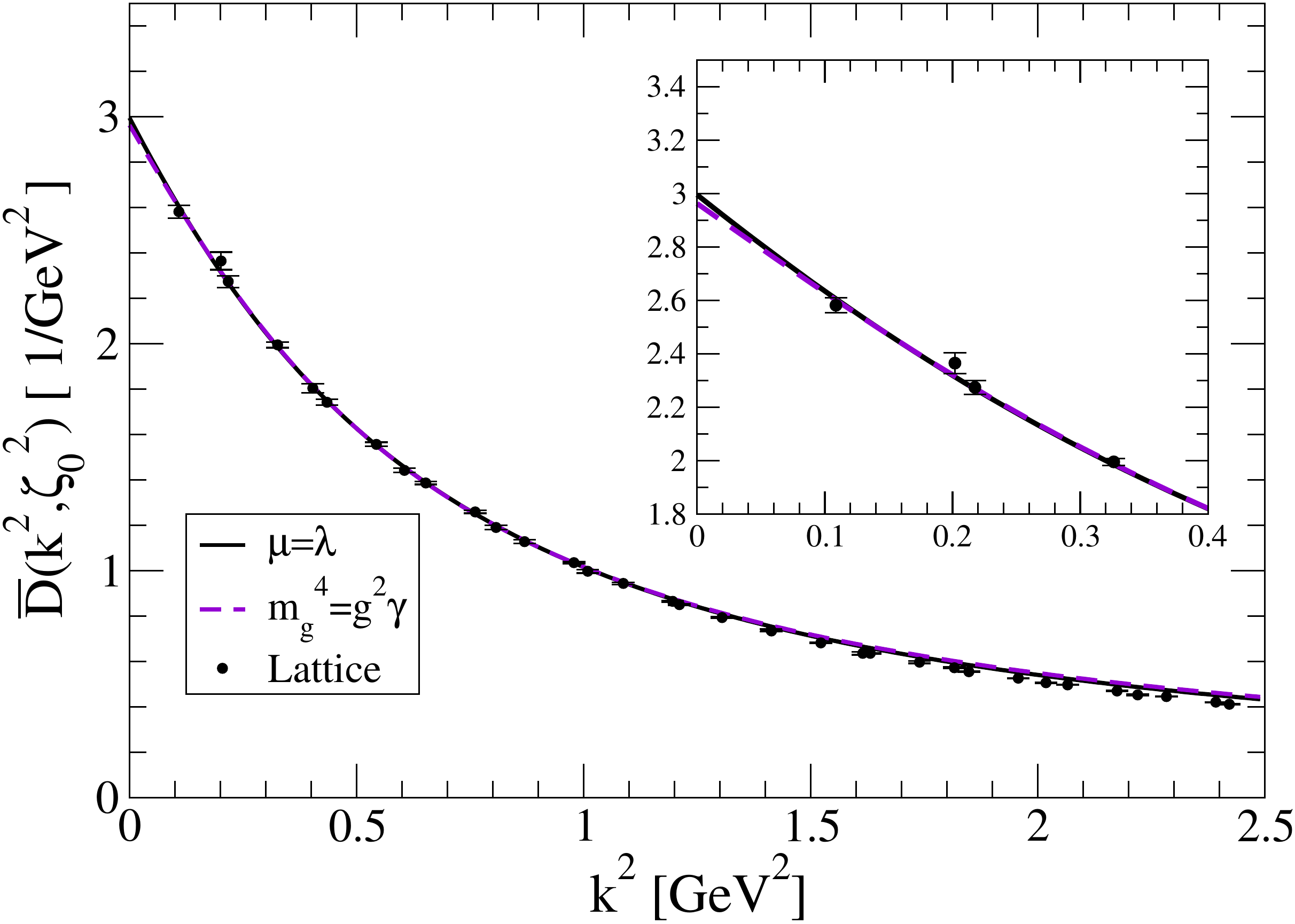}}
\caption{\label{equivalent}
\emph{Upper panel - (A)}.
Dot-dashed (purple) curve -- fit to quenched lattice results \cite{Bogolubsky:2009dc}, obtained as described in connection with Eqs.\,\eqref{allfit}, imposing Eq.\,\eqref{necessary} and $m_\gamma=m_g$.  (Best fit coefficients in Table.~\ref{latticefits}.)
Solid (black) curve, for comparison, the $\mu=\lambda$ curve from Fig.\,\ref{FitQuenched}.
\emph{Lower panel - (B)}.  Same as upper panel, but for unquenched results $(N_f=4)$ \cite{Ayala:2012pb}.}
\end{figure}

(\emph{i}) If $g^2\gamma = m_\gamma^4 \gg m_g^4$, then the Gribov horizon affects UV modes of the gluon.  The validity of standard perturbation theory shows this is not the case.  Consequently, $g^2\gamma \gg m_g^4$ is unrealistic: had it been favoured by lQCD results, then it would have been necessary to discard either or both those results and the horizon condition.

(\emph{ii}) The converse, $m_\gamma \ll m_g$, would have indicated that the gluon mass alone is sufficient to screen IR gluon modes, in which case the Gribov ambiguity could have no physical impact and any horizon term is redundant.

(\emph{iii}) Our analysis indicates that QCD occupies the middle ground: $m_\gamma \approx m_g$, with each of a size ($\sim 0.5\,$GeV) that one readily associates with emergent gauge-sector phenomena.  In this scenario, the gluon mass and the horizon scale play a nearly equal role in screening long-wavelength gluon modes, thereby dynamically eliminating Gribov ambiguities.  Moreover, together they set a confinement scale of roughly 1\,fm, identified with the location of the first zero in the configuration-space gluon two-point function.
We therefore conjecture that the gluon mass and horizon condition are equivalent emergent phenomena.
Plainly, unquenched lQCD results with better sensitivity to IR momenta are crucial before anything more can be said with certainty;  but if the quenched results are a reasonable guide, then such improvement would increase the likelihood that $m_\gamma = m_g$ is realised.

In the meantime, one can readily check whether $m_\gamma = m_g$ is consistent with available lQCD results; and, as apparent in Fig.\,\ref{equivalent}, that is certainly the case.  In this particular realisation of Scenario (\emph{iii}), the horizon term, and the $\langle A^2\rangle$ and ghost-field condensates may all be absorbed into a single running gluon mass, $m_g(k^2)$, whose dynamical appearance alone is then sufficient to eliminate the Gribov ambiguity and complete the definition of QCD.

\smallskip

\noindent\textbf{Acknowledgments}.
We are grateful for insightful comments from D.\,Binosi, L.\,Chang, D.~Dudal, C.\,Mezrag and J.\,Papavassiliou.
Research supported by:
National Natural Science Foundation of China, contract no.\,11435001;
National Key Basic Research Program of China, contract nos.\,G2013CB834400 and 2015CB856900;
U.S.\ Department of Energy, Office of Science, Office of Nuclear Physics, contract no.~DE-AC02-06CH11357;
and Spanish MEyC, under grant no.\,FPA2014-53631-C-2-P.
%




\end{document}